\documentclass[a4paper,english]{article}
\usepackage[T1]{fontenc}
\usepackage[latin1]{inputenc}
\usepackage{babel}
\usepackage{graphics}
\usepackage{german}
\usepackage{epsfig}

\makeatletter

\providecommand{\LyX}{L\kern-.1667em\lower.25em\hbox{Y}\kern-.125emX\@}

\makeatother
\begin{document}

\title{~
    \hspace*{\fill}{\small\sf http://xxx.lanl.gov/abs/hep-ph/0210178}\\
Towards the Finite Temperature Gluon Propagator in Landau Gauge Yang-Mills
Theory\thanks{
Talk given by A. Maas at the 40th International School of Subnuclear
Physics, August 29th-September 7th, Erice, Italy
}
}

\author{A.~Maas$^{1}$, B.~Gr\"uter$^{2}$, R.~Alkofer$^{2}$, J.~Wambach$^{1}$
\\\small{\em{${^1}$Institute for Nuclear Physics, Darmstadt University of Technology,}}\\\small{\em{Schlo{\ss}gartenstra{\ss}e 
9, D-64289 Darmstadt, Germany}}\\\small{\em{${^2}$Institute for Theoretical Physics, T\"ubingen University,}}\\\small{\em{ 
Auf der Morgenstelle 14, D-72076 T\"ubingen, Germany}}}

\date{}

\maketitle
\begin{abstract}
Yang-Mills theories undergo a deconfining phase transition at a critical
temperature. In lattice calculations the temporal Wilson loop and
\( Z_{3} \) order parameter show above this temperature a behavior
typical of deconfinement. A quantity of interest in the study of this
transition is the gluon propagator and its evolution with temperature.
This contribution describes the current status of an investigation
of the finite temperature gluon propagator in Landau gauge. It analyzes
the high temperature case. The resulting equations are compared to
the corresponding ones of three-dimensional Yang-Mills theory. Under
certain assumptions it is found that a kind of spatial {}``confinement''
is still present, even at very high temperatures.
\end{abstract}

\section{Introduction}

The gluon propagator at zero and finite temperature is of
many possible uses, although it is a gauge-dependent quantity. In
Landau gauge the ghost propagator is of equal importance.
Knowledge of these propagators can be used as input to phenomenological
calculations, e.g. in analysis of heavy ion collisions and the properties
of the produced state of matter. On the other hand these propagators
can be used to compare analytical calculations with lattice simulations
to estimate the effect of approximations and to gain insight into
the physical mechanisms underlying the lattice results and elucidate the dynamics of confinement.

This contribution describes the approach and the status of the calculation
of the finite-temperature gluon propagator in Landau gauge Yang-Mills (YM)
theory. The gluon propagator in the energy regime of interest for heavy
ion collisions is non-perturbative, and so must be the calculation.
As infrared singularities are anticipated, a continuum method is desirable.
Therefore Dyson-Schwinger equations are employed as a non-perturbative
approach. This will be described in section 2. The reason for choosing
this method are the encouraging results at vanishing temperature which will
be exemplified in section 3. Section 4 describes the necessary
extensions to access the equilibrium properties of gluons at finite temperature.
Section 5 will deal with the high-temperature limit and discuss
the comparison with a dimensionally reduced theory. Finally this contribution
ends with a summary and an outlook.

\section{Gluon and ghost propagator equations}

\subsection{Dyson-Schwinger equations}

\begin{figure}
\centerline{
\epsfig{file=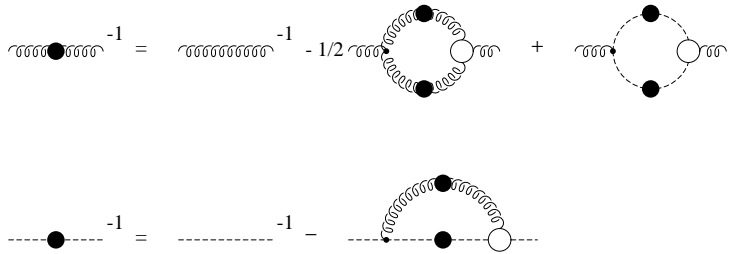,width=0.8\linewidth}
}
\caption{The truncated DSE used in the present work. 
The propagators with large full dots denote the
fully dressed propagators while the ones without are the perturbative
ones. A vertex denoted by a small full dot represents a perturbative vertex 
while the open circles indicate the fully dressed ones. In addition to
the employed truncation all vertex functions are approximated by bare ones.}
\label{fig1}
\end{figure}

Dyson-Schwinger equations \cite{Dyson:1949ha} (DSE) are a well-known tool for non-perturbative
calculations in several areas of physics. The DSE
for YM theories can be obtained by using functional derivatives
\cite{Alkofer:2000wg,Roberts:2000aa}. 
However, this approach produces an infinite hierarchy
of equations: each equation of a given \( n \)-point function depends on
higher \( n \)-point functions. Since it is not possible to solve
such an infinite coupled system, it is necessary to truncate it. There is
no a-priori information about which truncation is sufficient to keep
the relevant physics. The truncation scheme employed here is justified
by comparison of the results at zero temperature with lattice results,
which show that all qualitative features and many quantitative features
survive. The truncation scheme is to keep only the equations for the
2-point functions and also only up to three-point vertex level as
is depicted in figure \ref{fig1}.

In addition, it is assumed that multiplicative renormalization holds
also in the non-perturbative regime and that therefore wavefunction
renormalization and vertex renormalization can be cast into renormalization
constants \( Z_{i} \). Renormalization is performed using a momentum
subtraction scheme. The resulting equations after inserting the color structure and removing an overall factor of $\delta_{ab}$ are\footnote{All 
quantities are defined
using the conventions of ref.~\cite{Alkofer:2000wg}.} 
\begin{eqnarray}
D_{G}^{-1}(k)&=&-\widetilde{Z_{3}}k^{2}+\widetilde{Z_{1}}N_{c}\int 
\frac{d^{4}q}{(2\pi )^{4}}ik_{\mu }D_{\mu \nu }(k-q)G_{\nu }(q,k)D_{G}(q) \; ,
\\
D_{\mu \nu }^{-1}(k)&=&Z_{3}D_{\mu \nu }^{tl-1}(k)-\widetilde{Z_{1}}N_{c}\int 
\frac{d^{4}q}{(2\pi )^{4}}iq_{\mu }D_{G}(k+q)D_{G}(q)G_{\nu }(k+q,q)
\nonumber \\*
&&\qquad\qquad  +\frac{1}{2}Z_{1}N_{c}\int \frac{d^{4}q}{(2\pi )^{4}}
\Gamma _{\mu \rho \alpha }^{tl}(k,-(k+q),q) \nonumber \\*
&& \qquad \qquad\qquad \qquad D_{\alpha \beta }(q)D_{\rho \sigma }(k+q)
\Gamma _{\beta \sigma \mu }(-q,q+k,-k)\; .
\end{eqnarray}
To render the equations dimensionless, the dressing functions \( G \)
and \( Z \) are defined via their relation to the propagators
\begin{equation}
D_{G}(k)=-\frac{G(k)}{k^{2}}\; ,\quad
D_{\mu \nu }(k)=P_{\mu \nu }(k)\frac{Z(k)}{k^{2}}
\end{equation}
for the ghost and the gluon, respectively, and the transverse projector
is given by \( P_{\mu \nu }=\delta _{\mu \nu }-k_{\mu }k_{\nu }/k^{2} \).

It remains to fix the fully dressed vertices, since they are unknown
as long as the equations for the three point functions are discarded.
It is possible to construct these vertices using the Slavnov-Taylor
identities to constrain them as much as possible and this has been
done \cite{vonSmekal:1997is}. At zero temperature, explicit calculations show
that vertices constructed in this way only slightly improve the
result compared to perturbative vertices \cite{Alkofer:2000wg}. The effect
of other non-perturbatively dressed vertices has been studied in 
ref.~\cite{Lerche:2002ep}. Since constructed vertices induce a significant
complexity to the problem it is therefore more reasonable in this
first approach to finite temperature to keep only the perturbative
vertices. 
%These vertices follow the conventions used in \cite{Alkofer:2000wg}.

\subsection{Gauge symmetry}

To obtain only scalar equations it is useful to contract the gluon
equation with the projector \( P_{\mu \nu } \). However, it turns out that such
a contraction produces spurious quadratic divergencies. 
%These are
%terms which would lead to an ultraviolet divergence of the integrals
%for a constant dressing function. 
They stem from the fact that the
truncation of the DSE violates gauge symmetry.
To remove these divergencies and therefore the effects of gauge symmetry
violation it is possible to use instead of the transverse projector
\( P_{\mu \nu } \) other projectors which project onto states without
gauge symmetry violations. This can be accomplished e.g. by using a generalized projector
\begin{equation}
P_{\mu \nu }^{\zeta }=\delta _{\mu \nu }-\zeta \frac{k_{\mu }k_{\nu }}{k^{2}}
\end{equation}
where \( \zeta  \) is a real parameter. Choosing \( \zeta =d=4 \) removes
the spurious quadratic divergencies \cite{Brown:1988bm}. This procedure
is not unique and introduces ambiguities in the value of numerical
coefficients. This effect has been studied by varying the value of
\( \zeta  \) while removing manually the spurious divergencies. The
results at zero temperature demonstrate that these effects are small
\cite{Fischer:2002hn}.

A second item with respect to gauge symmetry in the non-perturbative
regime are Gribov copies. It has been shown that for the solutions
of DSE in Landau gauge it is sufficient to require
positive semi-definiteness of the dressing functions \( G \) and \( Z \) for all momenta 
to stay within the first Gribov horizon \cite{Zwanziger:2001kw}.
This is however not yet a full solution to the problem, since 
%recent investigations have shown that 
there are also Gribov copies within
the first Gribov horizon. It is  possible, in principle, to solve the
problem by either introducing a second set of ghost fields to fully
fix the gauge or by adding a new term to the DSE
\cite{Zwanziger:2001kw,Zwanziger:1993dh}. But, since lattice calculations indicate that the
influence of Gribov copies inside the first Gribov horizon is small
\cite{Cucchieri:1997ns}, these considerable complifications are neglected
for now.

\subsection{Kugo-Ojima confinement criterion}

When studying the finite-temperature gluon and ghost propagator one of the main 
goals to be achieved  is the question of the fate of confinement.
It is therefore necessary to find a criterion to test for the presence
of confinement. Confinement means the absence of gluons from the physical
spectrum. This is for example possible if they do not have a K\"allen-Lehmann
representation. Kugo and Ojima were able to construct a criterion
to test for such a realization of confinement by assuming that the
gluons carry BRST charge \cite{Kugo:gm}. The criterion can be cast
in a simple form in Landau gauge \cite{Kugo:1995km}: Does the euclidean
ghost propagator diverge stronger than a particle pole as the euclidean
four momentum approaches zero, i.e. does \( G \) diverge for \( k\rightarrow 0 \)?
To utilize this criterion and to simplify the calculations, the analysis will be performed in euclidean space.

\section{Results at zero temperature}

\begin{figure}
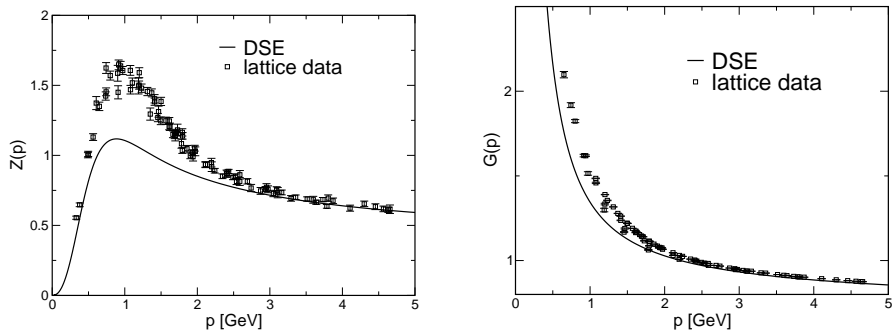

\begin{center}
 \epsfig{file=latt_gluecont2.eps,width=0.45\linewidth}
	\hspace{0.05\linewidth}
 \epsfig{file=latt_ghostcont2.eps,width=0.45\linewidth}
\end{center}
\caption{Comparison of lattice calculations \cite{Langfeld:2002dd} with results from 
DSE for \( N_{c}=2\). The left panel shows the dressing function 
of the gluon and the right panel the one for the ghost}
\label{fig2}
\end{figure}

The calculations at zero temperature have been carried out in 
\cite{Alkofer:2000wg,vonSmekal:1997is,Fischer:2002hn,Alkofer:2002ne}.
Figure \ref{fig2} shows a comparison of lattice calculations
with results of the DSE. These are calculations for \( N_{c}=2 \),
but the dressing functions have the same form for \( N_{c}=3 \), if 't Hooft
scaling is employed, since the DSE in this truncation
order depend only on the combination \( g^{2}N_{c} \).

Both propagators compare well with lattice results \cite{Langfeld:2002dd}, although
the gluon propagator misses some strength around 1 GeV.
It is likely that this is due to the neglect of the two-loop
diagram in the equations for the two-point functions~\cite{Alkofer:2000wg}.
The results show confinement according to the Kugo-Ojima criterion
with a divergence of \( G \) with an exponent of \( -\kappa  \)
in \( k^{2} \) with \( \kappa =0.5953 \). In addition, the gluon
dressing function vanishes at zero momentum with an exponent of \( 2\kappa  \)
\cite{Lerche:2002ep,Fischer:2002hn,Zwanziger:2001kw}. 
It is also possible to define a running coupling
constant and it turns out that according to these calculations there
is an infrared stable fixed point in YM theory. The good agreement
with lattice calculations and the fact that the Kugo-Ojima criterion is satisfied
motivates to use this method also at finite temperature.

\section{Extension to finite temperature}

\subsection{Formulation}

Since the primary interest are the equilibrium properties of gluons at finite
temperature, it
is convenient to use the Matsubara formalism. This induces that there
are now two independent variables, the discrete zero component of
the momentum and the absolute value of the three momentum. Additionally
there are now two possible tensor structures for the gluon propagator
\cite{Kapusta:tk} instead of only \( P_{\mu \nu } \). Both have to
be four dimensional-transverse due to gauge symmetry, but one is three-dimensional 
longitudinal and the other three-dimensional transverse.
They can be expressed as
\begin{eqnarray}
P_{T\mu \nu }&=&\delta _{\mu \nu }-\frac{k_{\mu }k_{\nu }}{k_{3}^{2}}+
\delta _{\mu 0}\frac{k_{0}k_{\nu }}{k_{3}^{2}}+\delta _{0\nu }\frac{k_{\mu }k_{0}}{k_{3}^{2}}
-\delta _{\mu 0}\delta _{0\nu }(1+\frac{k_{0}^{2}}{k_{3}^{2}})\nonumber \\
P_{L\mu \nu }&=&P_{\mu \nu }-P_{T\mu \nu }
\end{eqnarray}
where \( k_{3} \) denotes the magnitude of the three momentum and
the zero component of the momentum \( k_{0}=2\pi nT \) is the bosonic
Matsubara frequency. These projectors are orthogonal to each other
and satisfy \( P_{T\mu \mu }=2 \) and \( P_{L\mu \mu }=1 \). There
are therefore two independent dressing functions, \( Z_{L} \) and
\( Z_{T} \), and the gluon propagator is defined as
\begin{equation}
D_{\mu \nu }(k)=P_{T\mu \nu }\frac{Z_{T}(k)}{k^{2}}+P_{L\mu \nu }\frac{Z_{L}(k)}{k^{2}}
\; .
\end{equation}
By this definition \( Z_{L} \) and \( Z_{T} \) have to become equal
at zero temperature and equal to \( Z \).

It is possible to obtain two equations for \( Z_{L} \) and \( Z_{T} \)
by contracting the gluon DSE with \( P_{T\mu \nu } \)
and with \( P_{L\mu \nu } \) respectively. After inserting the perturbative
vertices, this results in the following set of equations for the three
dressing functions:
\begin{eqnarray}
\frac{1}{G(k)}&=&\widetilde{Z}_{3}+\widetilde{Z}_{1}\frac{g^{2}TN_{c}}
{(2\pi )^{2}}\sum_{n} \int d\theta dq_{3}
\nonumber \\
&& \qquad
(A_{L}(k,q)G(q)Z_{L}(k-q)+A_{T}(k,q)Z_{T}(k-q))\\
\frac{1}{Z_{L}(k)}&=&Z_{3}-\widetilde{Z}_{1}\frac{g^{2}TN_{c}}{(2\pi )^{2}}\sum_{n} \int d\theta dq_{3}P(k,q)G(q)G(k+q)
\nonumber \\
&+&Z_{1}\frac{g^{2}TN_{c}}{(2\pi )^{2}}\sum_{n} \int d\theta dq_{3}
\nonumber \\
&& \qquad (N_{L}(k,q)Z_{L}(q)Z_{L}(q+k)+N_{1}(k,q)Z_{L}(q)Z_{T}(q+k)
\nonumber \\
&& \qquad +N_{2}(k,q)Z_{L}(k+q)Z_{T}(q)+N_{T}(k,q)Z_{T}(q)Z_{T}(k+q))
\\
\frac{2}{Z_{T}(k)}&=&2Z_{3}-\widetilde{Z}_{1}\frac{g^{2}TN_{c}}{(2\pi )^{2}}\sum_{n} \int d\theta dq_{3}R(k,q)G(q)G(k+q)
\nonumber \\
&+&Z_{1}\frac{g^{2}TN_{c}}{(2\pi )^{2}}\sum_{n} \int d\theta dq_{3}
\nonumber \\
&& \qquad (M_{L}(k,q)Z_{L}(q)Z_{L}(q+k)+M_{1}(k,q)Z_{L}(q)Z_{T}(q+k)
\nonumber \\
&& \qquad +M_{2}(k,q)Z_{L}(k+q)Z_{T}(q)+M_{T}(k,q)Z_{T}(q)Z_{T}(k+q))
\end{eqnarray}
where \( T \) is the temperature and the sum runs over the bosonic Matsubara
frequencies of the gluon and the ghost. The integral kernels \( A_{i} \),
\( P \), \( R \), \( N_{i} \) and \( M_{i} \) are quite lengthy and will not
be quoted here. The trivial angular integration has been performed.

Since no additional divergencies arise at finite temperature \cite{Das:gg},
the renormalization procedure can be kept by employing temperature-independent
\( Z_{i} \).

\subsection{Gauge symmetry}

In the same sense as at zero temperature, spurious quadratic divergencies
arise due to the truncation and projection of the DSE. It is possible to remove 
such artifacts in the same, and
therefore also ambiguous, way as at zero temperature. Since there
are now two independent functions for the gluon dressing functions,
both projectors have to be modified and it is necessary to introduce
two real parameters. This can be accomplished using the following generalized
projectors
\begin{eqnarray}
P_{L\mu \nu }^{\xi }&=&\xi P_{L\mu \nu }+(1-\xi )(1+\frac{k_{0}^{2}}{k_{3}^{2}})
\delta _{\mu 0}\delta _{0\nu }
\nonumber \\
P_{T\mu \nu }^{\zeta }&=&\zeta P_{T\mu \nu }+(1-\zeta )(\delta _{\mu \nu }-
(1+\frac{k_{0}^{2}}{k_{3}^{2}})\delta _{\mu 0}\delta _{0\nu })\,. 
\end{eqnarray}
It turns out that all spurious divergencies are removed for \( \xi =0 \)
and \( \zeta =3 \). These values are interesting, since they force
the longitudinal projector to live only in the compactified dimension,
while the transverse part becomes a structure which is reminiscent
of the three-dimensional transverse projector.

\section{The high-temperature limit}

The finite-temperature regime is quite complicated and it is therefore
useful to start at some limiting cases and evolve their solutions
to finite temperatures. One possibility is to start at zero temperature
and to evolve the known solution to non-zero temperatures. This approach
is currently under investigation. The other approach is to start at
high temperatures. As this method provided already some insights it
will be described in the following.

\subsection{Definition of the high-temperature limit}

For the high-temperature limit, it is assumed that only the lowest
Matsubara frequency contributes. It is therefore described by first
letting the loop Matsubara frequency go to zero and then the exterior
frequency go to zero. The appearance of the explicit temperature dependence
can be removed by either measuring the momentum in units of temperature
or by defining a temperature dependent coupling constant \( g^{2}T=g_{3}^{2} \),
which is dimensionful and its usefulness will be discussed in the
next subsection.

By performing this limit, several remarkable observations can be made.
The first is, that many of the integral kernels vanish and that
the complete system becomes independent of \( \xi  \). Their exact
values are \goodbreak
$$
A_{L}=0,\:A_{T}=\frac{q_{3}^{2}\sin ^{3}\theta }{u_{-}^{2}},\:P=0,
$$
$$
N_{L}=0,\:N_{1}=-\frac{2q_{3}^{2}\sin ^{3}\theta }{u_{+}^{2}},\:N_{2}=
-\frac{2\sin ^{3}\theta }{u_{+}},\:N_{T}=0
$$
\begin{equation}
R=\frac{(q_{3}^{2}-(\zeta -1)w_{3}-\zeta q_{3}^{2}\cos ^{2}\theta )
\sin \theta }{k^{2}u_{+}}
\end{equation}
\[
M_{L}=\frac{((\zeta -1)k_{3}^{2}-4q_{3}^{2}+4(\zeta -1)w_{3}+
4\zeta q_{3}^{2}\cos ^{2}\theta )\sin \theta }{2k_{3}^{2}u_{+}},\:M_{1}=0,\:M_{2}=0\]
\[
M_{T}=\frac{\sin \theta }{2k_{3}^{2}u_{+}^{2}}((k_{3}^{2}+2q_{3}^{2})
((\zeta -9)k_{3}^{2}-4q_{3}^{2})+8(\zeta -3)w_{3}(k_{3}^{2}+q_{3}^{2})\]
\[
+((8\zeta q_{3}^{4}+(\zeta +7)k_{3}^{4}+4(5\zeta -1)k_{3}^{2}q_{3}^{2})+
4(4\zeta q_{3}^{2}+k_{3}^{2}(\zeta +3))w_{3}+4\zeta w_{3}^{2})\cos ^{2}\theta )\]
where \( u_{\pm }=k_{3}^{2}+q_{3}^{2}\pm 2k_{3}q_{3}\cos \theta  \)
and \( w_{3}=k_{3}q_{3}\cos \theta  \). Secondly, the three-dimensional
transverse projector now becomes the transverse projector of a theory
with only three dimensions, while the longitudinal projector only acts
in the compactified dimension. In addition, it is interesting
to note that the value of \( \zeta  \) coincides with that of the 
three-dimensional Brown-Pennigton projector.
But the most striking feature is, that if \( Z_{L}^{2} \) can be
neglected compared to \( Z_{T}^{2} \), then the equation for \( Z_{L} \)
decouples. This leads to an interesting comparison to the three-dimensional
theory.

\subsection{Comparison to three-dimensional Yang-Mills theory}

Compactified four-dimensional YM theory is not the same as
three-dimensional YM theory, since the zeroth component of
the four-dimensional gauge field becomes an additional Higgs field
\cite{Kajantie:1995dw}. However, lattice calculations indicate that this
Higgs field is unimportant and produces only a small effect on the
gluon propagator \cite{Cucchieri:2001tw}. It is therefore interesting to
compare to the three-dimensional YM theory only. The DSE
%Dyson-Schwinger equations 
can be calculated in the same way as in four dimensions
and yield 
\begin{eqnarray}
\frac{1}{G_{3d}(k_{3})}&=&\widetilde{Z}_{3}+\widetilde{Z}_{1}\frac{g^{2}_{3}
N_{c}}{(2\pi )^{2}}\int d\theta dq_{3}A_{3}(k_{3},q_{3})G_{3d}(q_{3})Z_{3d}
(k_{3}-q_{3})
\\
\frac{2}{Z_{3d}(k_{3})}&=&2Z_{3}-\widetilde{Z}_{1}\frac{g^{2}_{3}N_{c}}
{(2\pi )^{2}}\int d\theta dq_{3}P_{3}(k_{3},q_{3})G_{3d}(q_{3})G_{3d}
(k_{3}+q_{3})
\nonumber \\
&&+Z_{1}\frac{g^{2}_{3}N_{c}}{(2\pi )^{2}}\int d\theta dq_{3}(N_{3}(k_{3},
q_{3})Z_{3d}(q_{3})Z_{3d}(q_{3}+k_{3}))
\end{eqnarray}
with the integral kernels
\begin{eqnarray}
A_{3}&=&\frac{q_{3}^{2}\sin ^{3}\theta }{u_{-}^{2}},\:P_{3}=\frac{\sin \theta }
{2k_{3}^{2}u_{+}}(2(\zeta -1)w_{3}+q_{3}^{2}(\zeta -2+\zeta \cos 2\theta ))
\nonumber \\
N_{3}&=&\frac{\sin \theta }{2k_{3}^{2}u_{+}^{2}}((k_{3}^{2}+2q_{3}^{2})(k_{3}^{2}
(\zeta -9)-4q_{3}^{2}+8w_{3}(\zeta -3)(k_{3}^{2}+q_{3}^{2})
\nonumber \\
&& \qquad +((8\zeta q_{3}^{4}+(7+\zeta )k_{3}^{4}+4k_{3}^{2}q_{3}^{2}(5\zeta -1))
\nonumber \\
&& \qquad \qquad +4w_{3}
(4\zeta q_{3}^{2}+(\zeta +3)k_{3}^{2})+4\zeta w_{3}^{2})\cos ^{2}\theta )
\end{eqnarray}
using the three-dimensional Brown-Pennigton projector. It turns out,
that the resulting coupled equations for the three-dimensional ghost
and gluon are very similar. In fact, if \( Z_{L} \) can be neglected,
the three-dimensional coupling constant \( g^{2}_{3} \) is identified
with \( g^{2}T \) and the three-dimensional dressing function
\( Z_{3d} \) with \( Z_{T} \), then the above equations
become identical to the ones for three-dimensional YM
theory provided \( \zeta =3 \) is chosen. It is therefore tempting to identify
the longitudinal part of the four-dimensional gluon propagator with
the contribution from the Higgs field. This is supported by the fact
that the three-dimensional gluon propagator has also to be transverse
due to gauge symmetry and the three-dimensional longitudinal part
\( Z_{L} \) therefore cannot be part of it. This in turn would justify
the assumption that \( Z_{L} \) is indeed small and can be neglected
compared with \( Z_{T} \). However, only a detailed calculation of
these quantities can prove this assumption. Additional support
is found by the fact that in another approximation scheme, the Mandelstam
approximation, the same situation arises. Here is, however, no space
to detail these other calculations which have been done by us. Moreover,
recent lattice results support the finding of an infrared divergent
ghost dressing function at finite temperature \cite{Langfeld:2002dd} which
diverges weaker than at zero temperature as would be expected from
the above, since in three dimensions \( \kappa  \) is smaller \cite{Zwanziger:2001kw}.

If this turns out to be the case, it would have remarkable consequences,
since the three-dimensional YM theory indeed does confine
\cite{Zwanziger:2001kw}, and therefore a (spatial) kind of confinement would
still be present even at very high temperatures in YM theory.
This corresponds to lattice findings in that only the temporal Wilson
loop shows deconfinement while the Wilson loops in the not compactified
spatial dimensions do not \cite{Bali:1993tz}. This would underline the
fact that the magnetic sector of YM theory is non-trivial
even at high temperatures.

\section{Summary and outlook}

This contribution describes our recent progress in a calculation of
the finite-temperature gluon propagator in Landau-gauge YM
theory. It is a first step generalizing a self-consistent zero-temperature
formulation within the Matsubara formalism. Possibilities to circumvent
the problem of gauge symmetry violations due to the truncation of
the DSE have been discussed. A comparison of
the high-temperature equations of the compactified theory with the
corresponding equations for a lower-dimensional theory reveals interesting
properties. If the assumption of a negligible influence of the three-dimensional longitudinal
part of the gluon propagator at high temperature turns out to be correct,
then this proves the non-triviality of the magnetic sector of the YM
theory even at high temperatures. Note that this would imply the presence
of some confinement effects at very high temperatures. The physical
realization of such a spatial {}``confinement'' has still to be
understood. This may be possible as soon as the solutions of
the DSE are obtained. The corresponding calculations
are under way.

\newpage

\subsection*{Acknowledgments}

A.~M.\ thanks the organizers of the International School of Subnuclear
Physics for the very interesting meeting and for the opportunity to
present this talk. The authors are grateful to C. S. Fischer for helpful
discussions. 

This work is supported by the BMBF under grant number
06DA917 and by the European Graduate School Basel-T\"ubingen (DFG contract
GRK683).

\end{document}